# Cryoelectron Microscopy as a Functional Instrument for Systems Biology, Structural Analysis & Experimental Manipulations with Living Cells
*(A comprehensive review of the current works).*

Oleg V. Gradov
INEPCP RAS, Moscow, Russia
Email: o.v.gradov@gmail.com

Margaret A. Gradova
ICP RAS, Moscow, Russia
Email: m.a.gradova@gmail.com

*Abstract* — The aim of this paper is to give an introductory review of the cryoelectron microscopy as a complex data source for the most of the system biology branches, including the most perspective non-local approaches known as "localomics" and "dynamomics". A brief summary of various cryoelectron microscopy methods and corresponding system biological approaches is given in the text. The above classification can be considered as a useful framework for the primary comprehensions about cryoelectron microscopy aims and instrumental tools. We do not discuss any of these concepts in details, but merely point out that their methodological complexity follows only from the structure-functional complexity of biological systems which are investigated in this manner. We also postulate that one can employ some of the cryoelectron microscopic techniques not only for observation, but also for modification and structural refunctionalization of some biological and similar soft matter objects and microscopic samples. In other worlds, we start with the cryoelectron microscopy as a tool for the system biology and progress to its applying as an instrument for system biology and functional biomimetics; i.e. "system cryobiology" goes over into "synthetic cryobiology" or "cryogenic biomimetics". All these conclusions can be deduced from the most recent works of the latest years, including just submitted foreign papers. This article provides an up-to-date description of the conceptual basis for the novel view on the computational cryoelectron microscopy (in silico) approaches and the data mining principles which lie at the very foundation of modern structural analysis and reconstruction.

## I. TECHNICAL APPLICATIONS OF CRYOELECTRON MICROSCOPY

Since its development in early 1980s [31] cryo-electron microscopy has become one of the most functional research methods providing the study of physiological and biochemical changes in living matter at various hierarchical levels from single mammalian cell morphology [108] to nanostructures [69] and even the structure of biomacromolecules with a near-atomic resolution [19], i.e. at an all-atom configuration reconstruction [72]. In recent papers the cryo-electron microscopic techniques were reported to achieve the effective resolution up to several angstroms [12,71,130] and about 2-5 nm in a 3D representation [29]. To date the structural methods of cryo-electron microscopy have been involved in characterization of a number of supramolecular structures and complexes: from the well-studied fibrillar collagen structures [92], intraflagillar transport complexes of protozoa [91], tubulin in microtubules [125] and some other fibrillar and filamentous structures with the relative molecular weight of about gigadalton [101] up to such subtle elements of biological machinery as ribosomal complexes (e.g. those involved in cotranslational folding and translocation [62]), elements of secondary [95] and supersecondary protein structure [7], as well as the conformational effects in molecular mechanisms of their folding [25] and the mechanisms of expression in the complexes of encoding semantides [109] (in terms of Zuckerkandl and Pauling [133]). In addition to the conventional cryo-electron microscopy, modern structural analysis, particularly the determination of the protein secondary structure, requires the use of its precision 3D analog – cryo-electron tomography [9], which enables structure visualization at a wide scale / range of sizes, including the all-atom mapping [88] and a nanoscale cytophysiological ultrastructure analysis (so-called nanoimaging [63]. There is also an inherently close method of cryo-tomography – X-ray tomography with cooling under high pressure [123], which demonstrates approximately the same metrological characteristics.

## II. Algorithmic Tools for Cryo-Electron Microscopy

It is noteworthy that the ultra-high measurement accuracy requires the use of advanced mathematical apparatus and algorithmic processing: even the so-called direct methods of improving the image quality in cryo-electron microscopy are implemented only through the optimization of the number of summed and corresponding files with the unary cryo-electron micrographs [65,46]. This procedure requires a significant amount of computer time and robust high-performance mathematical processing systems. The modern trend in the hardware for such processing places a special emphasis on the CUDA technologies using video cards / graphics processing units (GPUs) providing the improvement of calculations [52], while for the purposes of data processing in cryo-electron microscopy in addition to the standard principles of machine learning in the pattern recognition [105,85] it is also useful to apply the following approaches:

1. flexible variants of the parameter fitting achieved by retraining during the search for extremes [3],

2. Bayesian probabilistic approaches in cryo-electron structural analysis [97],

3. algorithms of subtomogram analysis and in cryo-electron tomography [15,54],

4. eigen analysis with the search for eigenvectors in order to enable classification of the cryoelectron images upon their angular characteristics [106],

5. wavelet methods using multivariate functions and the Toeplitz structures [115], etc.

Except the above mentioned methods there are also a number of widely used conventional methods of structural and chemical profiling (e.g. "cryo-ultra-low-angle microtomy") [45] and other similar methods, underlying the CEDX (quantitative cryo-analytical scanning electron microscopy) approaches [78]. In other words, modern cryo-electron microscopic techniques are mostly automated data processing and interpretation methods, rather than data acquisition.

The above statement can be illustrated using the data provided by M. Stowell from the University of Colorado. It is seen from the Fig. 1 that even the similar structures with the equivalent morphometric parameters can demonstrate a high dispersion of geometrical characteristics, and in the case of different iterations of the mathematical and statistical data processing one can expect to obtain different morphologies and even different topologies (according to the number of connections) of the object under 3D reconstruction. The dynamics of the increase in accuracy of the measurements and morphological reconstructions from 1990 to 2000 was demonstrated by J. Frank in 2006 (see Fig. 2), which may be attributed to the improvement of the mathematical data processing methods. Thus, it is clear that the ambiguity of the structures' morphology after reconstruction in cryo-electron microscopy and tomography is characterized not only by quantitative (morphometric), but also by qualitative (geometric and topological) transformations.

Moreover, an ambiguous data interpretation in cryo-electron microscopic structure analysis necessitates the use of various mathematical models which determine the morphostructural representation of the result obtained, i.e. the above models may be one-to-one corresponded to the methods of data processing. Thus, there are several reports on cryo-electron microscopy using molecular dynamics simulations corresponding to flexible fitting methods [23]; pseudoatomic models corresponding to various approaches to the interpretation of full-atom structural analysis data on the angstrom resolution [84]; probabilistic 3D models of single particles, corresponding to the probabilistic (partly Bayesian) methods of data processing [35], etc. The simulation process should consider or derive *ab initio* the sample characteristics (namely, its solvation shell or a hydration layer [100]), as well as the noise simulation and the correction of the metrological readings drift [102] considering the process of image capturing [116]. With the increasing role of the model interpretation in cryo-EM data processing [76] there is a significant increase in the level of conjugation between modeling, docking (selection of the most favorable molecular orientation for the stable complex formation) and fitting of the full-atomic biomolecular structures in 3D mapping [4] (see Fig. 3), including the approaches and algorithms of the so-called cyclic or iterative refinement (see Fig. 4, 5) [127].

## III. CRYO-ELECTRON MICROSCOPY AS A TOOL FOR SYSTEMS BIOLOGY.

Cryo-electron microscopy, as well as cryo-electron tomography combine the principles of the computer systems-biological and structural-biological analyses *in silico*, *ex vivo* [81] and *in situ* [75]. Actually, among the numerous structural-biological «-omics», i.e. disciplines systematically studying individual classes of organic compounds in accordance with their biological role, cryo-electron microscopy and cryo-electron tomography can provide comprehensive 3D dynamical structural information about almost all of them, that in correlation microscopy is often called «localomics microscopy» and «dynamomics» (according to the classification of J. Parkinson). Such methods allow to study not only the different development phases or life cycles of individual biological objects[80], but also to perform the direct mapping of several characteristics of the molecule dynamics in molecular assemblies, visualized using the maximum-likelihood method [120], correlation microscopy for 3D structural analysis of the intra-cellular dynamic interactions using cryo-electron tomography [59] and a so-called "4D cryo-electron microscopy of proteins" [37]. Furthermore, in immunomics [16,112], considering the traditional use of a synchronous time-lapse cinematographic equipment in immuno-electron microscopy [8], it is reasonable to introduce dynamic cryo-electron measurements into cryo-immuno-electron microscopy [89], which at the same time will allow to avoid the image "blurring" artifacts by using the previous frames as the references for the reconstruction of the object shape and structure [18].

It would be also useful to consider some more examples illustrating the applicability of cryo-electron microscopy as a multivariate instrument in systems biology. First of all it concerns the investigations in membranomics and viromics. Ultrastructural studies of membranomes [103] using cryo-electron microscopy [70] (despite the lipid / phospholipid composition of the membranes [26], studied in the framework of classical lipidology) can not be directly attributed to the area of lipidomics or phospholipidomics [68,124], as well as lipoproteomics [53] (although it is possible to perform cryo-electron microscopic analysis of the lipoprotein structures [128] and membrane proteins [33]), since they predominantly consider the ultrastructural morphophysiological level of organization rather than the chemical composition of the sample. In addition to the applicability in membranomics, cryo-electon microscopy and tomography allow the study of membranous cytoplasmic organelles [50,28], including endosomes and autophagosomes involved in the recycling of the membrane material and its recyclization as a whole [32]. From the standpoint of computer lipidomics [38], concerning the dynamic changes in the membrane lipid content under signal transduction, of great significance is the problem of determination of the payload in the transport nanovesicles and clathrin-coated vesicles in the brain functional activity, successfully performed by cryo-electron tomography [51]. Current research in the field of system-biological and applied viromics [77] also can not be reduced to a common viral genomics [60] due to the main focus on the structural aspects of data interpre-tation in cryo-electron microscopy, tomography and similar conventional methods [22]. Emphasis is placed on the structure of the viral complexes with exogenous factors, such as the membrane protein DAF [127], conformational shift in a poliovirus antigen [74], capsid structural analysis [131], subvirus particles in membranes [44] and the entire virus particles [2], as well as on the studies of structural and biochemical affinity using specific selective affinity grids [61]. The above studies do not concern the genetic aspects of the viral structure, hence they can not be reduced to the structural genomics. In other words, there are specific areas of application for cryo-electron microscopy in systems biology, which fail to find a more suitable analytical tool than cryo-electron microscopy with further multiparametric (multivariative) data interpretation on a computer. Therefore, the complexity of many areas concerning the problem under investigation determines the complexity of the parameters studied.

## IV. MULTIPARAMETRIC CROSS-VERIFICATION AS A CONSEQUENCE OF THE CELLULAR COMPLEXITY.

To confirm the systems nature of the approach proposed it is necessary to demonstrate the compatibility of cryo-electron microscopy with the other methods of structural analysis in order to provide a comparative cross-validation of the results obtained [36]. Let us consider such an approach with reference to the comparative studies using cryo-electron and optical microscopy [14,129], hybridization of cryo-electron microscopy with fluorescent techniques [57,96] and Raman microspectroscopy with laser tweezers [110]. It is also noteworthy that cryo-electron microscopic data should be comparable with the results of the other fine methods of structural analysis, such as small-angle neutron scattering [11] and spectrochemical methods important in the determination of the molecular / supramolecular structure.

## V. CRYO-ELECTRON MICROSCOPY AS A TOOL FOR STRUCTURAL MICROMACHINING

The structural analysis of a biological sample at the electron microscope resolution provides its electron or ion beam micromachining *in situ*, as reported in a recent paper describing the focused ion beam micromachining of the eukaryotic cells (*Dictyostelium discoideum*) under cryo-electron tomography [93]. The major trend of this approach is the experimental scheme which allows a simultaneous imaging and micromachining [94], and the prospects of such hybridization make it possible to design molecular machines (mimetics for molecular cell machinery [34]) and perform various research manipulations with them using cryo-electron microscopy [27,39]. Such controlled micromanipulations become possible due to the fact that the registration of the particle beam-induced microscale structure mobility at a near-atomic resolution has yet been achieved in cryo-electron microscopy [73], providing a feedback between the registration systems and the controller of the microbeam impact. Neither cryo-electron microscopy itself, nor the similar methods based on it posses significant restrictions on the chemical composition of the samples. Both organic biological structures and biomineralized or biomimetic objects – magnetosomes [1], and even non-biological zwitterionic systems such as binary surfactants [86] can be successfully studied by cryo-electron microscopy. Therefore, there are also no restrictions on the composition of the structures formed under cryogenic electron microscopic micromachining.

## VI. THE ANALYTE DISTRIBUTION MAPPING AND THE DESCRIPTOR IMAGING IN PHENOMES.

A system-biological approach in cryo-electron microscopy together with the absence of the cryo-EM-micromachining and microstructuring product classification according to their chemical composition are expected to provide mapping and qualitative distribution of various analytes, both specific within the particular area of systems biology or "-omics" [5] and nonspecific. To date, numerous synthetic compounds of either biochemical or non-biological nature are used as imaging agents in cryo-electron microscopy [13,87], but even in pure biological and biomimetic research the main aim of the combined structural and chemical analysis is to perform non-destructive testing, i.e. ultrastructure visualization without any chemical modifiers and specific markers [43].

Meanwhile, the mapping procedure can be easily performed without any external imaging agents – according to the sample extinction parameters (electron or optical density) controlled by resolvometry with the quantitative data output in numerical form [64] using model validation [30]. Since the single values without considering the context digital information are insufficient from the metrological standpoint [20], the comparative mapping (which assumes several points as references in density) may be aimed at the analysis of the local sample electron density variations (and, hence, the image optical density) and distribution gradients. The corresponding data acquisition for some purposes (in particular, the protein structure analysis with cryo-EM maps using flexible fitting methods [90]) can be easily automated and partially delegated to the PC connected to the network servers with databases and libraries for the structure analysis and identification.

## VII. ARE THERE ANY ALTERNATIVES TO THE TARGET MAPPING & CONTRAST?

By analogy with the confocal microscopy which allows using a non-fluorescent imaging for the intracellular components in the absence of fluorescent labels, based on the intrinsic properties of the sample and the light beam (e.g., differential interference contrast), in cryo-electron microscopy there is a number of methods which allow the imaging of samples inappropriate for cryo-EM studies. One of them is Zernike phase contrast method which is used both in cryo-electron microscopy and cryo-electron tomography [41,118]. A similar result can be obtained using the frozen-hydrated biological samples with a Boersch phase plate [10,117]. It should be noted that Zernike phase contrast is achieved in a similar way in X-ray cryomicroscopy [121], so the microparticle visualization in both cases leads to mathematically similar results and can be subjected to computer processing using the algorithms similar to those applied for the confocal "tomographic" images of various scales in visible, ultraviolet and near-infrared spectral ranges [66,67]. In this case during the image processing one needs not only to eliminate the digital registration artifacts obvious in cryo-electron microscopy and tomography [107], but also to avoid the beam-induced blurring reducing the image contrast [83]. The image quality can be also improved by using the algorithms of the astigmatism and defocus estimation through the modulation transfer function - contrast transfer function (CTF) [114], based on the power spectra analysis with a quadrature filter, which is directly related to the nature of the phase contrast [132].

## VIII. FROM SYSTEMS TO SYNTHETIC BIOLOGY THROUGH CRYO-ELECTRON MICROMACHINING.

Various factors reducing the image clarity obviously reduce the analytical quality of the 3D cryo-EM images and the focused ion beam (FIB) positioning accuracy during the biological object modification under cryo-electron fixation, particularly in "cryo-FIB" technique with cryo-EM registration [119] and its quintessence – "cryo-FIB-SEM" [98]. The development of complex systems with a synchronized sample analysis and modification is expected within the modern tendency to the combination of the systems biology as an analytical approach with the reverse constructive approach – synthetic biology or the design of some arbitrary artificial biological / biomimetic structures. Thus, the analysis of natural biological structures (e.g., flagella and ciliary structures of the protozoa [17]) and semi-synthetic objects (those synthesized artificially from the biochemical components), such as 3D protein crystals for crystallographic studies [82], lays the foundation for cryo-electron 3D-morphological analysis of synthetic DNA-origami [6] and even more complex hybrid artificial structures, such as an artificial synaps [43]. Despite the fact that in some cases due to the planar 2D structure of the sample a less dimensional analysis may be sufficient enough to obtain the morphological data (cryo-electron microscopy instead of cryo-tomography) [58,126], the increase in the analysis dimension generally leads to the enlargement of the data amount, and hence, improves its heuristic value, providing new opportunities for the quantitative analysis [113].

## IX. THE PROSPECTS OF FURTHER DEVELOPMENT, SIMPLIFICATION & IMPLEMENTATION OF THE METHOD.

As a result of the above advantages the prevalence of cryo-electron microscopy in Europe, the USA, Japan and China, as well as the number of studies performed using cryo-EM has been continuously increasing in recent years, and most of the corresponding methods and setups developed at the end of the last century are considered now retrospectively [111].

Creating a system infrastructure for high-throughput high-resolution cryogenic electron microscopy [104], implementation of novel advanced analytical methods, such as self-pressurized rapid freezing (SPRF) technique [49]; design of portable plungers for intact cryo-electron microscopy sample preparation in natural environments [24] and development of hybrid instruments for complex structural analysis based on cryo-EM and cryo-tomographic techniques are among the most recent problems to be addressed in this area. A breakthrough in the above research was the development of correlative cryogenic X-ray tomography with cryo-light and electron microscopy [21]. As a result of the increasing infrastructure and technical availability of cryo-electron techniques there is a strong need in popularization of these methods for biologists, physicians and other specialists who are not directly engaged in working with cryogenic or cryoelectron microscopy. For this purpose there is a special kind of papers - "primers" [79] with specific recommendations on the organization of research work using combined methods of cell or tissue analysis and modification [55], individual molecular biological protocols for the analysis of particular compounds specific for different "-omics" by cryo-electron microscopy [42], guidelines for grid preparation for the sample fixation [48], based on the combinatorial testing of various sample compositions and preparations. Biological non-destructive analysis safe for both the sample and its environment using low dose techniques and cryo-electron microscopy has been also successfully performed [40].

## X. CONCLUSIONS.

Nowadays cryo-electron microscopy has become a usual method of biological research and ultrastructural control in bioengineering adapted for specific parameters under investigation. The already achieved image quality, spatial resolution and sensitivity is of applied character [99] and does not affect the physical, technical and engineering basis of the method. The introduction of additional axes, sources and data acquisition systems [47] to the basic construction is similar to the same process in optical microscopy, where the two-, three- and even four-objective multi-axis systems have recently emerged due to the progressive tendency of the image dimension increase providing a quantitative empowerment of the known method [56,122]. According to the private communications of the foreign colleagues, the new achievements in cryo-electron microscopy may be attributed to the variations of the sample environment, the chamber pressure, selection of useful combinations of detecting and imaging methods, the operation parameter optimization for the instrumentation and new robust algorithms of adaptive manipulation control. There is also a prospect of designing novel scanning cryogenic systems with the sweep output using alternative variables: cryoscopic analogs of Lorenzian magnetic microscopes capable of single magnetosome mapping; accelerator "synchro-cryo-micro-mass-spectrometers" which allow to determine the sample composition *in situ* during its structure analysis and / or micromanipulations with it; functional analogs of scanning thermal and thermoelectric microscopy based on the gradients of thermal and dynamic properties under low temperatures, allowing to distinguish between the living and non-living objects on a submicron scale and at the relatively short time intervals from the start of the sample preparation), etc. One of the most promising prospects of the aforementioned re-search performed mainly abroad is to develop cryobiology at a supramolecular level and to expand the range of problems in systems biology addressed within the electron microscopic cryobiology using the methods of multivariate biophysical screening.

The authors are grateful to their colleagues from Europe and the USA for helpful discussions, providing the full texts of their papers (including the unpublished ones) and for their kind permission to use original illustrations in our review.